\documentclass[12pt]{article}

\usepackage{newtxtext,newtxmath}
\usepackage{graphicx}

\usepackage[letterpaper,margin=1in]{geometry}

\renewenvironment{abstract}
	{\quotation}
	{\endquotation}

\date{}

\makeatletter
\renewcommand{\fnum@figure}{\textbf{Figure \thefigure}}
\renewcommand{\fnum@table}{\textbf{Table \thetable}}
\makeatother

\usepackage{scicite}

\usepackage{url}

\renewcommand{\dh}{\fontencoding{T1}\selectfont{\symbol{240}}}
\def\scititle{
	Quantum Fluids of Light in 2D Artificial Reconfigurable Aperiodic Crystals with Tailored Coupling
}
\title{\bfseries \boldmath \scititle}

\author{
	Sergey~Alyatkin$^{1\ast}$,
	Kirill~Sitnik$^{1}$,
	Valt\'{y}r~K\'{a}ri~Dan\'{i}elsson$^{2}$,
    Yaroslav~V.~Kartashov$^{3}$\and
    Julian~D.~T\"opfer$^{1}$,
    Helgi~Sigur{\dh}sson$^{2,4}$,
    Pavlos~G.~Lagoudakis$^{1\ast}$\and
	\small$^{1}$Hybrid Photonics Laboratory, Skolkovo Institute of Science and Technology,\and
    \small Territory of Innovation Center Skolkovo, Bolshoy Boulevard 30, building 1, Moscow \& 121205, Russia.\and
	\small$^{2}$Science Institute, University of Iceland, Dunhagi 3, Reykjav\'{i}k \& IS-107, Iceland.\and
    \small$^{3}$Institute of Spectroscopy of Russian Academy of Sciences, Fizicheskaya Str., 5, Moscow \& 108840, Russia.\and
    \small$^{4}$Institute of Experimental Physics, Faculty of Physics, University of Warsaw, \and
    \small ul.~Pasteura 5, Warsaw \& PL-02-093, Poland.\and
	\small$^\ast$Corresponding author. Email: S.Alyatkin@skoltech.ru; P.Lagoudakis@skoltech.ru\and
    }

\begin{document} 

% Insert the title and author list
\maketitle

% Abstract, in bold
% There are strict length limits, and not all formats have abstracts.
% Consult the journal instructions to authors for details.
% Do not cite any references in the abstract.
\begin{abstract} \bfseries \boldmath
Aperiodic crystals are the intermediates between strictly periodic crystalline matter and amorphous solids. The lack of translational symmetry combined with intrinsic long-range order endows aperiodic crystals with unique physical characteristics, while at the same time dramatically enriching the spectrum and localization properties. Here, we demonstrate exciton-polariton condensation in a two-dimensional Penrose tiling with $C_{10}$ rotational symmetry - the first signature of quasicrystalline order in a quantum fluid of light. We identify a regime, wherein near-perfect delocalization and synchronization of a quantum fluid of light occurs at mesoscopic length-scales extending beyond 100$\times$ the healing length and the size of each individual condensate. Realizing long-range order in fully reconfigurable aperiodic crystals of nonlinear, and open-dissipative quantum fluids, lays the foundations for testing a broad range of universality classes of continuous phase transitions beyond the limits of mathematically verifiable models in regular lattices.
\end{abstract}

% The first paragraph of any Science paper does NOT have a heading
% Nor is it indented
\noindent

Physical systems constructed upon aperiodic (or quasiperiodic) order are generally considered obscure and difficult to predict in comparison to strictly periodic structures~\cite{janot2012}. With the discovery of quasicrystals in 1982 by Schechtman et al., who observed five-fold rotational symmetry in the electron diffraction pattern of Al-Mn~\cite{Schechtman_1984PRL}, it became abundantly clear that aperiodic order can also form in the solid state. Ever since, scientists have artificially designed aperiodic structures in a wide variety of physical systems to gain insights into their properties in a well-controlled environment. As a result, Penrose tilings, Fibonacci chains, Sierpi\'{n}ski gaskets have been studied in photonic~\cite{Vardeny_NatPho2013, Xu_NatPho2021, Wang_NatPho2024} and electronic~\cite{Collins_NatComm2017, Kempkes_NatPhys2019} systems, with plasmon polaritons~\cite{Verre_ACSNano2014}, in thin-film ferromagnets~\cite{Watanabe_SciAdv2021}, ultracold atomic systems~\cite{Schreiber_Science2015, Viebahn_PRL2019}, and laterally modulated one-dimensional (1D) semiconductor optical cavities~\cite{Tanese_PRL2014, Baboux_PRB2017, Goblot_NatPhy2020}.

Current interest in the study of aperiodic structures is continuously fueled not only by unusual evolution of excitations in such physical platforms, but also by the discovery of advanced methods for constructing such objects with new types of symmetry. For instance, up to now, it was believed that a two-dimensional (2D) quasicrystal can be tiled with at least two distinct shapes of tiles, the prototypical example being the P3 Penrose quasicrystal made up of a pair of thin and thick rhombi~\cite{penrose1974role}. However, a recently mathematically discovered form of tiling - an aperiodic monotile requiring only one type of a tile to build up the entire quasicrystal~\cite{Smith_Arxiv2023}. Not realized physically yet, such system is predicted to have spectral similarities to graphene, including six-fold symmetry and Dirac-like features~\cite{Monotile_PRL}. 

The long-range order of 2D quasicrystals is a consequence of their self-similarity, which results in a fractal reciprocal lattice manifested in the Bragg peaks of diffracted waves that underline an ordered scattering mechanism. In this regard, artificial photonic quasicrystals offer a unique advantage to explore such intricate reciprocal patterns through far field measurements~\cite{Man_Nature2005, Vardeny_NatPho2013, Che_PRL2021}. However, given the photon's weak interaction strength, the focus has mostly been on linear (single particle) dynamics such as wave transport~\cite{Matsui_Nature2007}, localization~\cite{Wang_NatPho2024} including Anderson localization of light~\cite{Segev2013}, topology~\cite{Bandres_PRX2016} and relation to higher dimensional physics~\cite{Kraus_NatPhy2016} and nearly exclusively in conservative quasicrystal systems. In such systems it was shown that linear quasicrystals impose unconventional localization properties for wave excitations \cite{Wang_NatPho2024}, dramatically different from localization properties in other periodic or aperiodic media~\cite{Wang2019}. Up-to-date little is known about the optically nonlinear~\cite{Freedman_Nature2006, Freedman_NatMat2007} or lasing~\cite{Notomi_PRL2004, Vitiello_NatComm2014} properties of quasicrystals, and even less when these two paramount features are combined together in a dissipative environment. We anticipate that the study of such physical systems will open a principal route to a wide range of novel wave-propagation and localization phenomena that may acquire completely unexpected features.% in a open-dissipative environment. %To the best of our knowledge, reconfigurable quasicrystals remain elusive in open-dissipative systems. 

In this article, we demonstrate nonequilibrium Bose-Einstein condensation of ballistic microcavity exciton-polaritons, or ``polariton lasing'', in a 2D Penrose quasicrystal. We utilise a structured optical pumping that creates expanding polariton condensates at the vertices of the quasicrystal tiles. We observe the emergence of long-range quasiperiodic order in the extended polariton system evidenced by the formation of multiple sharp Bragg peaks in the condensate photoluminescence (PL) displaying the characteristic ten-fold rotational symmetry for the Penrose tiling. The build up of long-range coherence in extended aperiodic lattices is made possible through the ballistic flow of polaritons that couple distant condensate neighbours due to the strong polariton-polariton interactions. % in aperiod and the implemented structured optical pumping technique that leads to ballistic polariton flows that couple between distant condensate neighbours in aperiodic potentials. 
As a consequence, coherent polariton waves from each condensate can undergo multiple scattering processes in the quasicrystal, probing its structure at a much larger scale than in evanescent (tightly-bound) lattices. Our results offer the first glimpse into 2D quasicrystalline exciton-polariton systems characterized by long-range ballistic coupling, strong nonlinearities, and large coherence lengths.

\subsection*{Experimental realization of 2D Penrose tiling}

\begin{figure} % Do NOT use \begin{figure*}
	\centering
	\includegraphics[width=0.9\textwidth]{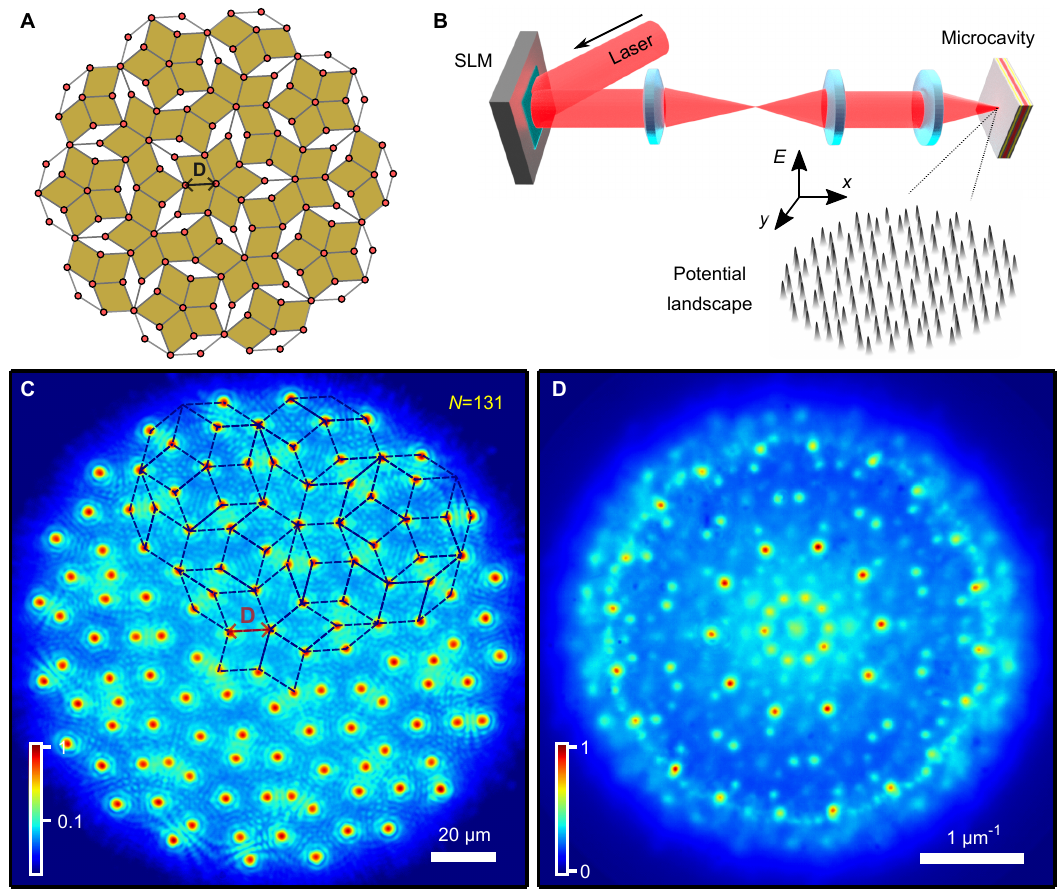} % for an image file named example_figure.*
	% Pick an appropriate width - in print, figures are usually one or two columns wide, which can
	% be approximated by 0.3\textwidth or 0.6\textwidth respectively. Use appropriate label sizes.

	% Captions go below figures
	\caption{\textbf{Realization of 2D polariton quasicrystal based on the Penrose tiling.}
		($\textbf{A}$) Fragment of the Penrose mosaic made of thin and thick rhombi with side length $D$. ($\textbf{B}$) Sketch of the experimental setup to imprint an aperiodic tiling on semiconductor microcavity. The non-resonant excitation laser is shaped with a spatial light modulator (SLM) in aperiodic arrangement of Gaussian beams [red circles in (A)], defining the potential and pump landscape felt by polaritons. ($\textbf{C}$) Measured real-space photoluminescence (PL) at $P=1.4 P_\mathrm{thr}$ of the Penrose tiling (partially shown to guide the eye) with the number of vertices $N=131$ and $D=13.2$~$\upmu$m. ($\textbf{D}$) Corresponding reciprocal-space PL features $C_{10}$ rotational symmetry with narrow Bragg peaks indicating formation of coherent state in the aperiodic tiling.}
	\label{Fig1} % give each figure a logical label name
\end{figure}

To implement the P3 Penrose tiling, schematically shown in Fig.~\ref{Fig1}A, we use all-optical lattice imprinting on a planar GaAs-based microcavity with embedded InGaAs quantum wells~\cite{Cilibrizzi_APL2014}. Figure~\ref{Fig1}B schematically shows the experimental setup. To transform a nonresonant pulsed Ti:Sapphire laser emission (pulse-width $\uptau\approx$~5~ps) into an ordered array of Gaussian beams forming the Penrose tiling, we use a phase-only spatial light modulator (SLM). Using a modified Gerchberg–Saxton (GS) algorithm~\cite{Gerchberg_1972}, we calculate the SLM phase mask and utilize an active feedback loop~\cite{Toepfer_Optica} to create the desired excitation pattern, consisting of pumping spots with equal intensities. This approach results in a uniform PL intensity distribution when the sample is pumped near the polariton condensation threshold ($P=P_\mathrm{thr}$).

Figure~\ref{Fig1}C shows experimentally measured real-space polariton PL at pump power $P=1.4 P_\mathrm{thr}$  for the Penrose tiling with the number of vertices $N=131$ and rhombus side length set to $D=13.2$~$\upmu$m. Formation of macroscopic coherent state and the phase-locking between occupied  vertices of the Penrose tiling is manifested in the interference fringes between the condensates. This macroscopic quantum state formed in dissipative system is very different from e.g. cold atoms that are tightly bound in quasiperiodic optical lattices~\cite{Schreiber_Science2015, Viebahn_PRL2019, Zhu_PRA2024}. The typical polariton blueshifts (potential amplitude) coming from the tightly focused excitation spots is around $V_0 \approx 2$~meV whereas the recoil energy is around $E_r = \uppi^2 \hbar^2/(2mD^2) \approx 0.05$~meV for a typical polariton mass of $m \approx 5.2 \times 10^{-5} m_0$, where $m_0$ is the free electron mass. This large energy contrast $V_0/E_r \gg 1$ underlines that the condensate dynamics is determined by slowly decaying propagating waves instead of evanescent coupling between strongly localized modes on individual optical potential minima that have been studied in other condensed matter and photonic platforms~\cite{Schreiber_Science2015, Tanese_PRL2014, Baboux_PRB2017, Kempkes_NatPhys2019, Collins_NatComm2017, Viebahn_PRL2019, Goblot_NatPhy2020, Xu_NatPho2021, Wang_NatPho2024}.

As the pump power exceeds the condensation threshold, the reciprocal-space PL reveals formation of many organized Bragg peaks following the fractal composition of the quasicrystal's reciprocal lattice vectors, 
\begin{equation}
    \mathbf{K} = \sum_{i=1}^5 n_i \mathbf{k}_i, \qquad n_i \in \mathbb{Z}.
\end{equation}
\noindent\ Here, $\mathbf{k}_i = b [\cos{(\uppi i/5)}, \sin{(\uppi i/5)}]$ where $b=(2\uppi/D)(2 \varphi^2/5)$ and $\varphi=(1+\sqrt{5})/2$ is the golden ratio~\cite{janot2012, Notomi_PRL2004}. Moreover, the relative integrated intensity $I_\mathrm{rel}=I_\mathrm{peaks}/I_\mathrm{total}$ of these Bragg peaks, where $I_\mathrm{total}$ is the total integrated reciprocal-space PL, changes with pump power and reaches a maximum value at $P=1.4 P_\mathrm{thr}$ [see Materials and Methods~\cite{methods} for details]. Here, $I_\mathrm{rel}$ serves as a measure of the contrast between the coherent polariton signal and the incoherent background. This power-dependent optimal coherence is a feature of optical polariton lattices where ballistic coupling is the strongest against dephasing effects from the background photoexcited exciton reservoir~\cite{Toepfer_Optica}. The reciprocal-space PL at such ``optimized excitation conditions'' in Fig.~\ref{Fig1}D features $C_{10}$ rotational symmetry - a clear signature of the quasicrystalline order formation. We note here that the observed synchronization area exceeds by 100$\times$ the healing length $\upxi$ of each individual condensate, estimated at $\upxi\approx1.3$~$\upmu$m.  

The observation of synchronization in aperiodic structures made of gain-localized nonlinear phase-amplitude oscillators (polariton condensates) is a fundamentally novel phenomenon, clearly different from the observations in periodic lattices of coupled condensates~\cite{Masumoto2012,Schneider2016,Whittaker_PRL,Ohadi_PRB2018,Toepfer_Optica}. It is known that each pair of spatially separated condensates can synchronize in-phase, out-of-phase (antiphase), or occupy a nonstationary oscillatory state (limit cycle) caused by mode competition between polariton standing waves in the cavity plane, that depends strongly on their separation distance~\cite{Toepfer_time-delay, Alyatkin_PRL}. This implies that the node-to-node distance and the in-plane momentum of polaritons are critically important parameters for condensate synchronization~\cite{Ohadi_PRX}. From this point of view, tuning the excitation parameters and the lattice geometry allows for observation of robust single-mode polariton lasing in periodic lattices when the wavenumber of outflowing condensate polaritons matches a reciprocal lattice number~\cite{Alyatkin_NatComm, Alyatkin_APL2024}. In contrast, the aperiodic Penrose mosaic, although it possesses self-similarity despite the lack of translational symmetry, consists of thick and thin rhombuses wherein each node is surrounded by several neighbors with different and incommensurate separation distances~\cite{janot2012}. The natural consequence of this is a fractal band structure with multiple energy states occupied by polaritons as observed also in one-dimensional Fibonacci chains~\cite{Tanese_PRL2014} (see  fig.~\ref{fig:sup5} for numerically resolved Penrose quasicrystal dispersion). Surprisingly, the multiple aperiodically scattered (from pump spots) polariton waves support synchronization of distant condensates, forming a single-mode macroscopic state with well-defined phase (Fig.~\ref{Fig1}C,D). These findings underpin the efficiency of ballistic polaritons to couple through multiple available diffraction orders on the optical quasicrystal's isofrequency surface in a driven-dissipative system (see fig.~\ref{fig:sup1}).       

\subsection*{Build-up of coherence and quasicrystalline order}

An apparent question that rises is how many vertices the tiling must consist of to have in its spectrum all the main features inherent to infinitely extended aperiodic tiling. To address this issue, we examine the build-up of the quasicrystalline order as the system grows in size, i.e. when the number of tiles increases. We fix the spatial spacing between the central vertices of the tiling at $D=13.2$~$\upmu$m (like in Fig.~\ref{Fig1}C) and incrementally expand the aperiodic structure by adding the vertices at the periphery. Figure~\ref{Fig2}A shows experimentally measured real-space polariton PL for the Penrose tiling, consisting of only $N=46$ nodes. Similar to observations for $N=131$ nodes, for pump power $P\geq P_\mathrm{thr}$ the Bragg peaks appear in reciprocal space. We find that their relative integrated intensity $I_\mathrm{rel}$ is maximized at $P=1.43 P_\mathrm{thr}$. For this pump power and low number of vertices ($N=46$) we still could observe signatures of $C_{10}$ rotational symmetry in Fig.~\ref{Fig2}B, hinting on the long-range order formation. However, the Bragg peaks corresponding to such spatially limited tiling were found to be broader, overlapping more with incoherent emission, pointing out to stronger ``inelastic'' scattering of polaritons.   

Further on, we seek to identify the minimum number of vertices $N$, where $C_{10}$ rotational symmetry becomes the dominant feature. %We further extend the Penrose tiling by increasing the number of vertices $N$. 
The measured real- and reciprocal-space polariton PL for $N=86$, $N=111$, $N=151$ are shown in Fig.~\ref{Fig2}C-D, Fig.~\ref{Fig2}E-F and Fig.~\ref{Fig2}G-H, respectively. We stress that for each of the imprinted structures we varied the pump power from below threshold to $P\approx1.6 P_\mathrm{thr}$ and extracted the intensity distributions of polariton PL for the pump powers which maximize the relative intensity of the Bragg peaks $I_\mathrm{rel}$. In other words, for each $N$ we maximize the amplitude of the mutual coherence function for coupled polariton condensates. It is evident from Fig.~\ref{Fig2}I, that for fixed spacing $D=13.2$~$\upmu$m, the pump conditions satisfying optimal coherence correspond to practically the same value $P\approx1.4 P_\mathrm{thr}$ (see red and yellow circles for $N=151$ and $N=46$, respectively).

Figure~\ref{Fig2}J shows obtained dependence of the maximum value of $I_\mathrm{rel}$ of the Bragg peaks (for a given rhombus side length $D$) as a function of number of vertices $N$. One can clearly see a saturation of the Bragg peak contrast above the number of vertices $N\geq$~110; see the dashed curve plotted in Fig.~\ref{Fig2}J to guide the eye. Here, we note that the measured maximum values of $I_\textrm{rel}$ for $N=111, 131$ and 151 have reached saturation. We draw the reader's attention to similarity of the reciprocal-space PL distributions shown in Fig.~\ref{Fig2}F, Fig.~\ref{Fig1}D and Fig.~\ref{Fig2}H. Therefore, we conclude that synthetic polariton analogue of the Penrose quasicrystal is effectively formed for the number of vertices exceeding $N\geq$~110 in our cavity, as marked schematically with a light green area in Fig.~\ref{Fig2}J. This point is defined by the coherence length of the ballistic polariton condensates which for regular lattices is around $\sim 10^2$~$\upmu$m~\cite{Toepfer_Optica}. Indeed, the diameter of the Penrose lattice, i.e. maximum distance between two condensates, for $N=111$ vertices is $L \approx 129$~$\upmu$m which means that the lattice size now exceeds the coherence length leading to saturation of the Bragg peak contrast. In the Supplementary text we provide an analysis of the linear eigenmodes of quasicrystal pump landscape for different number of vertices $N$, as shown in fig.~\ref{fig:sup7} and fig.~\ref{fig:sup8}.   

\subsection*{Probing localization properties of polariton quasicrystals}

Observation of the quasicrystalline order in Fig.~\ref{Fig2} would not be possible in the Penrose tiling without ballistic in-plane propagation of polaritons, which implies a spatial delocalization of the wavefunction. A common approach to characterize the degree of the state localization relies on the calculation of the inverse participation ratio ($\textrm{IPR}$) parameter:  
\begin{equation}
\textrm{IPR}=\frac{\int d \mathbf{r}|\psi(\mathbf{r})|^4}{\left(\int d \mathbf{r}|\psi(\mathbf{r})|^2\right)^2}.
\end{equation}

\noindent\ For eigenstates of finite-size systems of spatial dimension $L$ it is usually possible to introduce the scaling relation $\textrm{IPR}\propto$$L^{\gamma}$, where $\gamma=0$ corresponds to the localized states, while $\gamma=-2$ corresponds to spatially extended (delocalized) states. Therefore, in order to probe the properties of the polariton wavefunction in aperiodic Penrose tiling we extract the $\textrm{IPR}$ as a function of the pump power $P$ and size $L$ (defined as largest distance between vertices in quasicrystal) of the aperiodic polariton structure like in Fig.~\ref{Fig1}, scaled with number of vertices $N$ for a fixed spacing $D$. For this, in Fig.~\ref{Fig3}A we plot $\log{(\textrm{IPR})}$ versus $\log{(L/D)}$ dependence calculated from the measured real-space PL for the pump power in the range from $P=P_\mathrm{thr}$ to $P=1.5 P_\mathrm{thr}$. Using the linear fits of the experimental data, we extract the slope coefficient of the curves at different pump power, which equals the scaling factor $\gamma$. Obtained values of $\gamma$ as a function of pump power $P$ are given in Fig.~\ref{Fig3}B. Our analysis confirms a high degree of polariton wavefunction delocalization with $\gamma=-1.76\pm0.09$ [corresponding to critical states~\cite{Goblot_NatPhy2020, Zhu_PRA2024}] at $P=P_\mathrm{thr}$ and $\gamma=-1.98\pm0.11$ at $P=1.2 P_\mathrm{thr}$, approaching the value of $\gamma=-2$, characteristic of delocalized states (see green dashed line). These experimental results are fully supported by mean field numerical simulations (see fig.~\ref{fig:sup3}).
The main feature of quasicrystals is the long-range order conventionally evidenced by scattering measurements. The intense Bragg peaks shown in Fig.~\ref{Fig2} are clearly distinct, due to effective scattering of polaritons on the pump-induced aperiodic potential. Taking into account our experimental proof of the quasicrystalline order present in polariton Penrose tiling, we further test the robustness of the synchronized aperiodic array of condensates to defects - vacancies of vertices. For this, we utilize the uniqueness of our imprinted system, namely the ability to control individual pump spots and condensates. By this, we can artificially introduce a vacancy-defect in an aperiodic array, which may affect the scattering of polaritons. Below we describe our observations for the Penrose tiling with the number of vertices $N=131$ and the rhombus side length $D=13.5$~$\upmu$m, where we gradually increased the number of defects (absent vertices) and probed the system behavior. 

Figures~\ref{Fig4}A,B show real- and reciprocal-space PL for initial ``ideal'' disorder-free aperiodic structure. The results are given for the optimum pump power (maximizing the value of $I_\mathrm{rel}$), determined in the same manner as for the results in Fig.~\ref{Fig2}. Analysis of the experimental data confirms that polaritons indeed condense into a single energy state, as confirmed by the measured spectrum power scan in Fig.~\ref{Fig4}C. Next, using a random number generator we have removed the vertices with pointed numbers (5 out of 131), and repeated the power scan measurements for the Penrose tiling with 5 artificially created vacancies. The results are shown in Fig.~\ref{Fig4}D-F. The real-space PL in Fig.~\ref{Fig4}D still reveals clear interference fringes between the vertices. As it appears, injected polaritons still acquire sufficient momentum as to coherently couple (see momentum distribution in Fig.~\ref{Fig4}E), while scattering across multiple nodes despite introduced defects to the aperiodic tiling. Slightly increased finite incoherent background in reciprocal-space PL distribution in Fig.~\ref{Fig4}E still does not preclude from observation of expected system of Bragg peaks with $C_{10}$ rotational symmetry. We note effective broadening of the emission linewidth, visible from the spectrum in Fig.~\ref{Fig4}F at $P\geq1.3 P_\mathrm{thr}$, accompanied by weak satellite lower energy state hardly visible for initial disorder-free system.

To follow the changes in the Penrose quasicrystals we sequentially implement and characterize the tilings with number of defects set to $N_\mathrm{vac}=2, 7, 10, 14$ and finally to $N_\mathrm{vac}=21$. Surprisingly, even in the structure with 21 vacancies (16$\%$ of not occupied vertices), the polariton condensates still efficiently phase-lock and reveal a quasicrystalline order as shown in Fig.~\ref{Fig4}G-H. Clearly, the contribution from inelastic scattering across (full of defects) tiling becomes visible and results in modification of the spectrum at $P\geq1.3 P_\mathrm{thr}$. However, for the pump power in the range from $P=P_\mathrm{thr}$ to $P=1.2P_\mathrm{thr}$ polariton PL is mono-mode within the spectral resolution of our setup ($\approx20$~$\upmu$eV), similar to our  observations for the Penrose tiling with the number of vacancies $0\leq N_\mathrm{vac}\leq20$. Figure~\ref{Fig4}J shows the extracted average width of the Bragg peaks as a function of number of vacancies. Analysis reveals minor monotonous broadening of the peaks in reciprocal space, demonstrating the robustness of the quasicrystal to artificially introduced disorder in sense of persistence of the long-range order in the system (see also fig.~\ref{fig:sup4} for corresponding mean field solutions).

\subsection*{Reconfigurable polariton quasicrystals}
Finally, we explore how the rhombuses side length $D$, i.e. the characteristic spacing between pump spots, of the Penrose tiling affects the synchronization of polariton condensates. For this we fix the number of vertices ($N=106$) in the excitation pattern and vary $D$ as shown in Fig.~\ref{Fig5}. We find that as we decrease the rhombus side length from $D=13.5$~$\upmu$m (Fig.~\ref{Fig5}A) to $D=11.1$~$\upmu$m (Fig.~\ref{Fig5}C) the width of the Bragg peaks increases as visible from the normalized surface plots in Fig.~\ref{Fig5}B and Fig.~\ref{Fig5}D. It should be stressed that presented data corresponds to the pump power minimizing the width of the Bragg peaks for the given $D$ value. Increased width for smaller $D$ values can be attributed to strong inter-particle interactions and increased condensate overlap with the background incoherent exciton reservoir, which contributes to polariton dephasing. Nevertheless, the reciprocal-space PL in Fig.~\ref{Fig5}D still clearly demonstrates a signature of the quasicrystalline order.

Next, we implement the Penrose tiling with $D=10.1$~$\upmu$m, as shown in Fig.~\ref{Fig5}E. In contrast to more expanded tiling in Fig.~\ref{Fig5}A, the system in Fig.~\ref{Fig5}E with denser vertices does not reveal pronounced interference fringes. Moreover, the contribution of the PL intensity from the areas in between the vertices has increased. As a consequence, the reciprocal-space PL in Fig.~\ref{Fig5}F does not display any distinct Bragg peaks intrinsic to the Penrose quasicrystal. Instead, one can see an individual peak at $k\approx0$ surrounded by the incoherent background. 

Further on, we set the rhombus length to $D=7.8$~$\upmu$m and find a complete destruction of the quasicrystalline order both in real (Fig.~\ref{Fig5}G) and reciprocal space (Fig.~\ref{Fig5}H). Strong repulsion of polaritons from the pumped vertices leads to smearing of the density distribution in real space. For any pump power above the threshold we did not observe synchronized condensates at the vertices that would exist at a single energy state. This is in a good agreement with previous observations in periodic lattices with small lattice constant, where the competition between higher and lower energy modes leads to multi-mode condensation with temporal density beating and smeared PL distribution (once averaged)~\cite{Alyatkin_NatComm}. The reciprocal-space PL in Fig.~\ref{Fig5}H shows non-uniform ring-like distribution with a tendency of shrinking towards zero-momentum $k=0$ as the pump power increases. From the experiments above, we conclude that the formation of the Penrose polariton quasicrystals is possible only for spacing $D\geq11.0$~$\upmu$m, allowing for efficient scattering of polaritons away from the vertices and their in-plane coupling, leading to mono-mode polariton lasing.

\subsection*{Discussion and outlook}

We demonstrated a 2D Penrose quasicrystal of ballistically coupled exciton-polariton condensates and investigated the coherence of the system on the number of vertices, as well as the robustness of the long-range order in the presence of artificially induced defects. Unlike periodic structures, where macroscopic particle coherence is routinely achieved through the fine tuning of the lattice constant~\cite{Toepfer_Optica,Tao2022_XYperovsk}, phase-locking in aperiodic arrays is complicated by the unique local environment of the vertices. The advantage of the investigated platform is its self-probing nature through the in-plane scattering of polaritons from the repulsive aperiodic potentials. This is in contrast to photonic waveguides arrays and structured microcavities, where the evanescent coupling limits the diversity of the excited states. Our approach opens new directions for the study of many-body physics in aperiodic potentials and synchronization phenomena in novel physical systems such as the Ulam's spiral, the Girih tiling, and the recently discovered monotile quasicrystal~\cite{Smith_Arxiv2023}.

%///////////////////////////////////////////

% Research Articles and Reviews split the text into sections using headings
% Use a short (up 6 words) descriptive phrase, not generic 'Results' or 'Conclusions'
% Most other formats do not have headings, see the journal instructions to authors for details

% If your text is very short you might need to uncomment the following line to avoid
% layout problems with the figures and tables.
%\newpage

%%%%%%%%%%%%%%%% MAIN TEXT FIGURES %%%%%%%%%%%%%%%

\begin{figure} % Do NOT use \begin{figure*}
	\centering
	\includegraphics[width=0.9\textwidth]{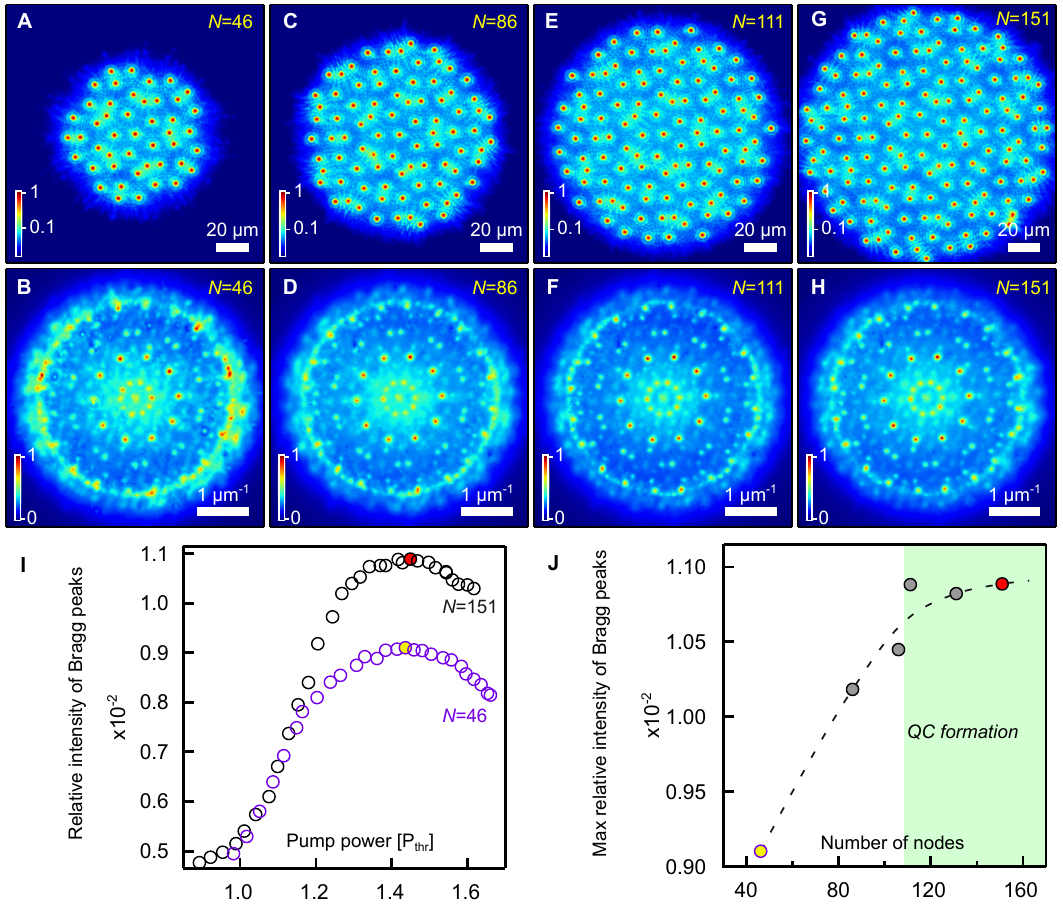} % for an image file named example_figure.*
	% Pick an appropriate width - in print, figures are usually one or two columns wide, which can
	% be approximated by 0.3\textwidth or 0.6\textwidth respectively. Use appropriate label sizes.

	% Captions go below figures
	\caption{\textbf{Formation of 2D polariton quasicrystal based on the Penrose tiling with rhombus side length $D=13.2$~$\upmu$m.}
		As the system scales up from $N=46$ vertices ($\textbf{A}$ and $\textbf{B}$) to $N=151$ ($\textbf{G}$ and $\textbf{H}$) the measured reciprocal-space PL in ($\textbf{B}$),($\textbf{D}$),($\textbf{F}$),($\textbf{H}$) reveals more pronounced Bragg peaks, confirming $C_{10}$ rotational symmetry. Corresponding real-space PL for $N=46, 86, 111, 151$ is shown in ($\textbf{A}$),($\textbf{C}$),($\textbf{E}$),($\textbf{G}$). Panel ($\textbf{I}$) shows integrated intensity of the Bragg peaks normalized to the total reciprocal-space PL intensity as a function of pump power for $N=46$ (purple circles) and $N=151$ (black circles). Clear maxima of the curves at $P\approx1.4 P_\mathrm{thr}$ correspond to the most coherent state. ($\textbf{J}$) Extracted maximum values of the relative Bragg peaks intensities as a function of number of vertices $N$ reveals saturation at $N\geq$110, see dashed line to guide the eye.}
	\label{Fig2} % give each figure a logical label name
\end{figure}

\begin{figure} % Do NOT use \begin{figure*}
	\centering
	\includegraphics[width=0.6\textwidth]{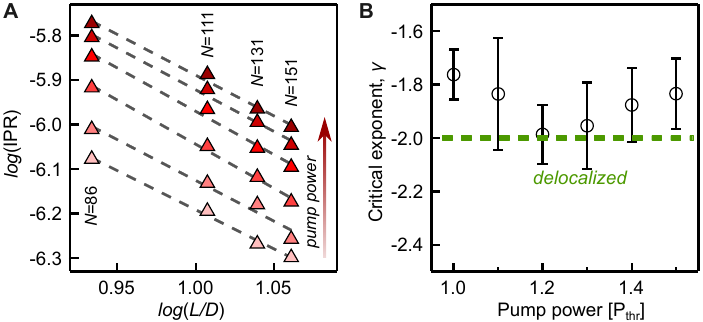} % for an image file named example_figure.*
	% Pick an appropriate width - in print, figures are usually one or two columns wide, which can
	% be approximated by 0.3\textwidth or 0.6\textwidth respectively. Use appropriate label sizes.

	% Captions go below figures
	\caption{\textbf{Experimental probing the localization properties of the polariton wavefunction in Penrose tiling shown in Fig. 2.}
		($\textbf{A}$) Experimentally extracted dependence of $\textrm{log}(\textrm{IPR})$ versus $\textrm{log(L/D)}$ for increasing pump from $P=P_\mathrm{thr}$ to $P=1.5 P_\mathrm{thr}$, where the $\textrm{IPR}$ is the inverse participation ratio, $L$ - system size and $D$ - rhombus side length. The slope of the linear fitting curves (dashed lines) corresponds to the scaling exponent $\gamma$, given in ($\textbf{B}$) as a function of pump power. Above the condensation threshold the wavefunction is highly delocalized ($\gamma=-1.98\pm0.11$ at $P=1.2 P_\mathrm{thr}$) due to ballistic in-plane propagation of polaritons with high momentum.}
	\label{Fig3} % give each figure a logical label name
\end{figure}

\begin{figure} % Do NOT use \begin{figure*}
	\centering
	\includegraphics[width=0.9\textwidth]{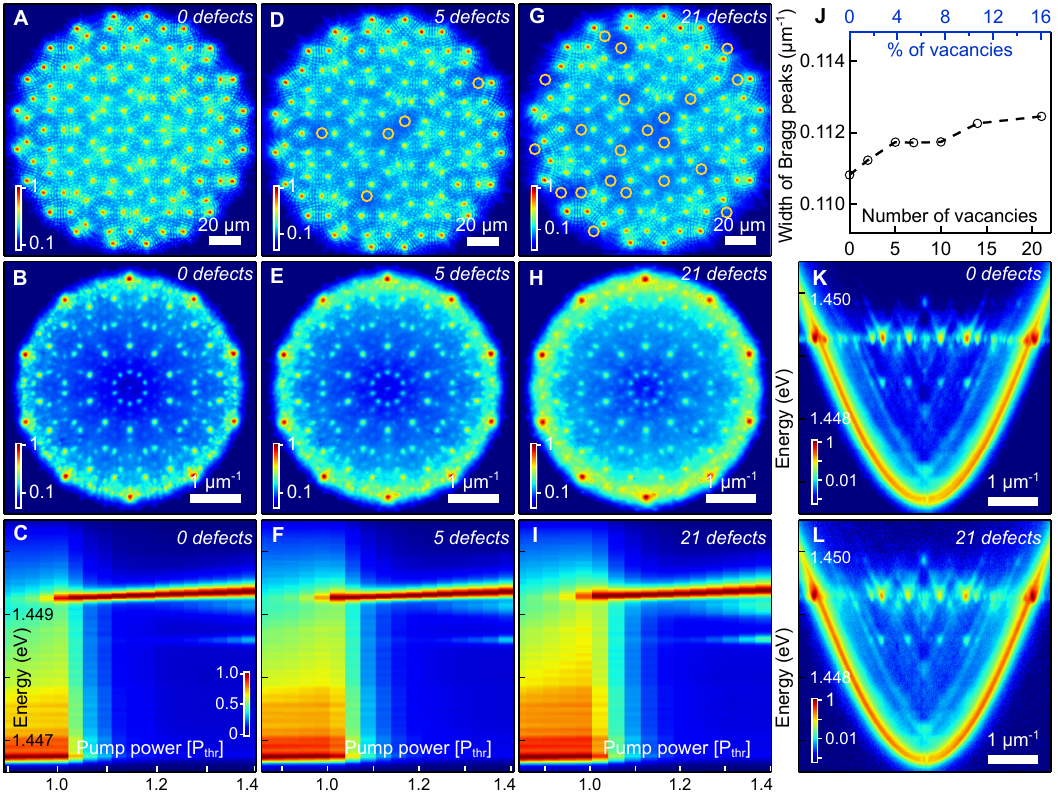} % for an image file named example_figure.*
	% Pick an appropriate width - in print, figures are usually one or two columns wide, which can
	% be approximated by 0.3\textwidth or 0.6\textwidth respectively. Use appropriate label sizes.

	% Captions go below figures
	\caption{\textbf{Long-range order coherence is preserved in quasicrystalline system even with artificially induced disorder - vacancies.}
		($\textbf{A}$),($\textbf{B}$),($\textbf{C}$) show real- , reciprocal-space PL and spectrum as a function of pump power for disorder-free (0 defects) quasicrystal. ($\textbf{D}$)-($\textbf{F}$) and ($\textbf{G}$)-($\textbf{I}$) correspond to the structures with 5 and 21 defects, respectively. Even with 21 (out of 131) unoccupied vertices the reciprocal-space PL in ($\textbf{H}$) features $C_{10}$ rotational symmetry inherent to quasicrystals, however, an incoherent background has increased as visible from (I). ($\textbf{J}$) The width of the Bragg peaks as a function of number of vacancies. ($\textbf{K}$),($\textbf{L}$) Normalized energy-resolved momentum-space PL for the Penrose tiling shown in (A) and (G) correspondingly, look almost indistinguishable.}
	\label{Fig4} % give each figure a logical label name
\end{figure}

\begin{figure} % Do NOT use \begin{figure*}
	\centering
	\includegraphics[width=0.9\textwidth]{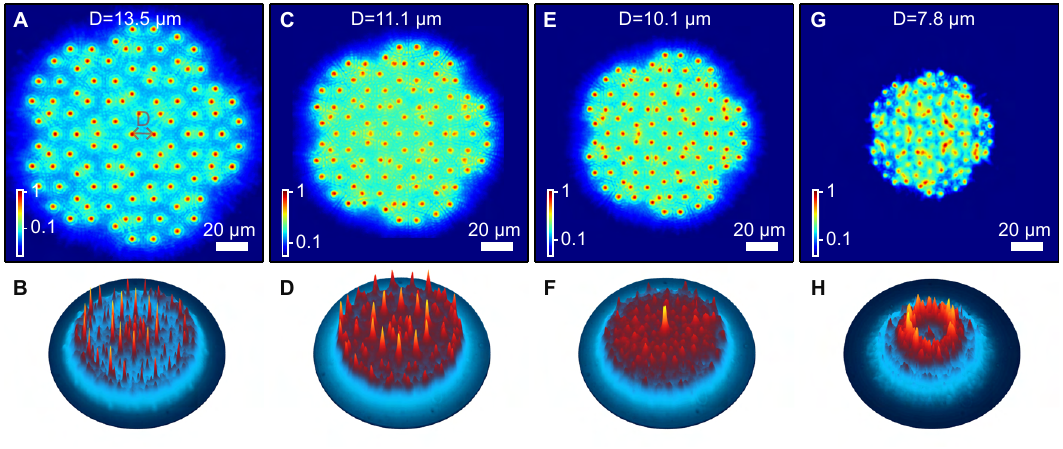} % for an image file named example_figure.*
	% Pick an appropriate width - in print, figures are usually one or two columns wide, which can
	% be approximated by 0.3\textwidth or 0.6\textwidth respectively. Use appropriate label sizes.

	% Captions go below figures
	\caption{\textbf{Control on transport properties (spectrum) of polariton quasicrystal achieved with a homogeneous structure compression for the fixed number of vertices $N=106$.}
		($\textbf{A}$ and $\textbf{B}$) Normalized real- and reciprocal-space PL for the spacing $D=13.5$~$\upmu$m. ($\textbf{C}$) Compression to $D=11.1$~$\upmu$m results in visible broadening of the Bragg peaks in ($\textbf{D}$). At even smaller $D=10.1$~$\upmu$m the polariton PL in between the nodes becomes more pronounced ($\textbf{E}$) with appearance of a single peak accompanied by strong background in momentum space in ($\textbf{F}$) and complete loss of $C_{10}$ rotational symmetry. ($\textbf{G}$) At $D=7.8$~$\upmu$m above condensation threshold polariton PL becomes smeared due to particles repulsion and their partial trapping outside the pumped vertices, visible in reciprocal-space ($\textbf{H}$).}
	\label{Fig5} % give each figure a logical label name
\end{figure}
%%%%%%%%%%%%%%%% REFERENCES %%%%%%%%%%%%%%%

\clearpage % Clear all remaining figures and tables then start a new page

% The list of references goes after the main text and before the acknowledgements
% When preparing an initial submission, we recommend you use BibTeX, like this:
%
\bibliography{Quasicrystals_Science} % for a file named science_template.bib
\bibliographystyle{sciencemag}

% After the paper has completed peer review and been revised ready for acceptance,
% you should comment out the lines above and copy-paste the contents of your .bbl
% file here instead. This will help ensure that our conversion software works correctly.
% Remember to re-run BibTeX first - check the timestamp!
%
% Example of the first three entries copy-pasted from science_template.bbl:
%
%\begin{thebibliography}{1}
%
%\bibitem{example}
%A.~N. {Author}, An example reference. \emph{Journal of Improbable Research}
%  \textbf{1}, 67 (2020).
%
%\bibitem{example2}
%F.~M. {Surname}, S.~{Author}, A second example. \emph{Interesting Research
%  Letters} \textbf{32}, 897 (2019).
%
%\bibitem{example_preprint}
%P.~{One}, P.~{Two}, P.~{Three}, {An unpublished preprint}. \emph{preprint}
%  (2021), arXiv:2101.12345.
%
%\end{thebibliography}

%%%%%%%%%%%%%%%% ACKNOWLEDGEMENTS %%%%%%%%%%%%%%%

\section*{Acknowledgments}
S.A. acknowledges Prof. Nikolay Gippius and Dr. Ekaterina Iashina for fruitful discussions.
\paragraph*{Funding:}
This study was supported by the Russian Science Foundation (RSF) (Grant No. 24-72-10118), https://rscf.ru/en/project/24-72-10118/. V.K.D acknowledges the Icelandic Research Fund (Rann\'{i}s), grant No. 239552-051. H.S. acknowledges the project No. 2022/45/P/ST3/00467 co-funded by the Polish National Science Centre and the European Union Framework Programme for Research and Innovation Horizon 2020 under the Marie Sk\l{}odowska-Curie grant agreement No. 945339.
\paragraph*{Author contributions:}
S.A. and K.S. carried out the experiments, S.A. analyzed experimental data, J.D.T. wrote a software for the data acquisition, V.K.D and H.S. performed theoretical modeling, Y.V.K. performed analysis of the modes structure, P.G.L. supervised the project, S.A. and H.S. wrote draft of the manuscript, all authors contributed to discussion and writing the manuscript.
\paragraph*{Competing interests:}
There are no competing interests to declare.
\paragraph*{Data and materials availability:}
All data are available in the main text or the supplementary materials.

%%%%%%%%%%%%%%%% SUPPLEMENT LIST %%%%%%%%%%%%%%%

% List the contents of your Supplementary Materials, including the numbers of any
% supplementary figures, tables, external data files etc. and any references that are
% cited only in the supplement. In this example, refs. 7-8 are cited only in the supplement.
% Fill out your numbers accordingly and delete any lines that aren't applicable.
\subsection*{Supplementary materials}
Materials and Methods\\
Supplementary Text\\
Figs. S1 to S8\\
References\\ %\textit{(7-\arabic{enumiv})}\\ % automatically fills out the last reference number
% (filling out the other numbers automatically is possible but fiddly and liable to break)

%%%%%%%%%%%%%%%% END OF MAIN TEXT %%%%%%%%%%%%%%%

\newpage

%%%%%%%%%%%%%%%% START OF SUPPLEMENT %%%%%%%%%%%%%%%

% Figures, tables, equations and pages in the supplement are numbered S1, S2 etc.
\renewcommand{\thefigure}{S\arabic{figure}}
\renewcommand{\thetable}{S\arabic{table}}
\renewcommand{\theequation}{S\arabic{equation}}
\renewcommand{\thepage}{S\arabic{page}}
\setcounter{figure}{0}
\setcounter{table}{0}
\setcounter{equation}{0}
\setcounter{page}{1} % not 0 as \newpage already started a supplementary page
% References continue the numbering from the main text.

%%%%%%%%%%%%%%%% SUPPLEMENT TITLE PAGE %%%%%%%%%%%%%%%

\begin{center}
\section*{Supplementary Materials for\\ \scititle}

% Author list for the supplement
% Indicate the corresponding authors, but do NOT include institutions here
% It would be nice if the template auto-generated this, but doing so is complicated...
Sergey~Alyatkin$^{1\ast}$,
	Kirill~Sitnik$^{1}$,
	Valt\'{y}r~K\'{a}ri~Dan\'{i}elsson$^{2}$,
    Yaroslav~V.~Kartashov$^{3}$,
    Julian~D.~T\"opfer$^{1}$,
    Helgi~Sigur{\dh}sson$^{2,4}$,
    Pavlos~G.~Lagoudakis$^{1\ast}$\\ % we're not in a \author{} environment this time, so use \\ for a new line
\small$^\ast$Corresponding author. Email: S.Alyatkin@skoltech.ru; P.Lagoudakis@skoltech.ru\\
\end{center}

% Fill out the numbers for each type of supplementary material,
% and delete any lines that aren't applicable.
% These are just example numbers that don't match the rest of this template.
\subsubsection*{This PDF file includes:}
Materials and Methods\\
Supplementary Text\\
Figures S1 to S8\\

\newpage

%%%%%%%%%%%%%%%% MATERIALS AND METHODS %%%%%%%%%%%%%%%

\subsection*{Materials and Methods}

The sample is cooled down to 4~K using a closed-cycle helium cryostat. The non-resonant laser is tuned at the first Bragg minimum of the microcavity reflectivity stop-band (1.5578~eV) to improve excitation efficiency and avoid heating of the sample. The laser radiation at fundamental repetition frequency of $\approx80$~MHz is additionally chopped using acousto-optical modulator at frequency of 5~kHz with a duty cycle of 3$\%$ to realize pulse train excitation. This ensures stable set temperature of the sample even for extremely large number of vertices in the tiling. Such excitation is used in all time-integrated experiments for real- and reciprocal-space PL measurements. The exciton-photon detuning is set to a negative value of $\updelta=-4$~meV to decrease effective mass of polaritons and facilitate their in-plane propagation and coupling~\cite{Toepfer_Optica, Alyatkin_PRL} between the pumped vertices of the aperiodic structure.

In order to extract the dependence of the Bragg peaks relative intensity $I_\mathrm{rel}$ on the pump power, shown in Fig.~\ref{Fig2}I, we analyze the corresponding reciprocal-space polariton PL intensity distributions. For this, at the given pump power we apply a mask capturing only the Bragg peaks with a diameter of 12~pixels (1 pixel $\approx$ 0.00855~$\upmu$m$^{-1}$) around each bright peak and calculate the integrated intensity of the ``masked'' pattern $I_\mathrm{peaks}$. Then at the same pump power using a big circular mask with a radius of $\approx$ 2.75~$\upmu$m$^{-1}$ capturing the whole reciprocal-space PL, we calculate the total integrated intensity $I_\mathrm{total}$ within the ``masked'' area. Next, we calculate the relative integrated intensity using $I_\mathrm{rel}=I_\mathrm{peaks}/I_\mathrm{total}$ and repeat the same for all pump power values. Then we extract the maxima of the obtained curves for different number of vertices in the Penrose tiling and plot this dependence as Fig.~\ref{Fig2}J.

%%%%%%%%%%%%%%%% SUPPLEMENTARY TEXT %%%%%%%%%%%%%%%

\subsection*{Supplementary Text}
\subsection*{S1 Simulating the time evolution of a 2D exciton-polariton condensate}

The scalar dynamics of the lower branch of an exciton-polariton condensate optically pumped by a nonresonant CW pump-profile $P(\mathbf{r})$ can be described with in the mean field approximation, resulting in a generalised Gross-Pitaevskii equation for the condensate wave-function, $\psi(\mathbf{r}, t)$, coupled to an exciton reservoir density, $n_R(\mathbf{r}, t)$.\cite{PhysRevB.77.115340}
\begin{align}
i\partial_t \psi &= \left[-\frac{\hbar}{2m}\nabla^2 + \alpha |\psi|^2 + G\left(n_R + \frac{\eta P}{\Gamma}\right) + \frac{i(Rn_R - \gamma_{LP})}{2}\right]\psi\label{gp} \\
    \partial_t n_R &= -(\Gamma + R|\psi|^2)n_R + P\label{nr}
\end{align}
Here, $m$ is the effective mass of the lower polariton branch, $\alpha$ is the self-coupling strength of the polariton condensate, $G$ is the coupling strength between the polaritons and excitons, $\Gamma$ is the exciton decay rate, $\gamma_{LP}$ is the decay rate of the lower polaritons, $R$ is the rate of stimulated scattering of polaritons into the condensate from the exciton reservoir, and $\eta$ is a phenomenological constant accounting for additional blueshift due to charge carriers and high-momentum exciton background.

The method used here to simulate the condensate-reservoir system is a split-step Fourier method.
Define $V(\mathbf{r}) = \alpha|\psi|^2 + G\left(n_R + \frac{\eta P}{\Gamma} + \frac{i(Rn_R - \gamma_{LP}}{2}\right)$. Then one can write Eq.~\eqref{gp} as 
\begin{equation}
    i\partial_t \psi = \left(\frac{-\hbar}{2m}\nabla^2 + V(\mathbf{r})\right)\psi 
\end{equation}
which, if the non-linear term is not too large, is well approximated by the formula
\begin{align}
    \psi(\mathbf{r}, t) = e^{(i\hbar\nabla^2/(2m) - iV(\mathbf{r}))t}\psi(\mathbf{r}, 0) 
\end{align}
for sufficiently small $t$.

If $t$ is not large, the propagator $e^{(i\hbar\nabla^2/(2m) - iV(\mathbf{r}))t}$ can be approximated by $e^{-i\frac{t}{2}V(\mathbf{r})}e^{i\hbar t\nabla^2 / (2m)}e^{-i\frac{t}{2}V(\mathbf{r})}$ according to the Baker-Hausdorff-Campbell formula. These operators are each diagonal in either $\mathbf{k}$-space or $\mathbf{r}$-space. Denoting the Fourier transform w.r.t.~position of a function $f(\mathbf{r}, t)$ by $\mathcal{F}\{f(\mathbf{r}, t)\}(\mathbf{k}, t)$, and the inverse by $\mathcal{F}^{-1}\{f(\mathbf{k}, t)\}(\mathbf{r}, t)$, a time-step by $\Delta t$ can be calculated with the formula
\begin{align}
    \psi(\mathbf{r}, t_0+\Delta t) \approx e^{-i\frac{\Delta t}{2}V(\mathbf{r})}\mathcal{F}^{-1}\left\{e^{-\frac{i\hbar \Delta t\mathbf{k}^2}{2m}}\mathcal{F}\left\{e^{-i\frac{\Delta t}{2}V(\mathbf{r})}\psi(\mathbf{r}, t_0)\right\}(\mathbf{k}, t_0)\right\}(\mathbf{r}, t_0). \tag{S5}
\end{align}
This method allows leveraging highly efficient fast Fourier transform algorithms, notably GPU accelerated algorithms, for numerical simulation of equation~\eqref{gp}.
The time evolution of the exciton reservoir is subsequently approximated by the formula
\begin{align}
    n_R(\mathbf{r}, t_0 + \Delta t) \approx \exp(-(\Gamma + R|\psi(\mathbf{r}, t_0 + \Delta t)|^2)\Delta t)n_R(\mathbf{r}, t_0) + P(\mathbf{r})\Delta t 
\end{align}

In the simulations shown in Figs.~\ref{fig:sup1}, \ref{fig:sup2}, \ref{fig:sup4} and \ref{fig:sup6} the parameters used are $\Gamma = 0.1$~ps$^{-1}$, $\gamma_{LP} = 0.2$~ps$^{-1}$, $\alpha = 0.0004$~$\upmu$m$^2$ps$^{-1}$, $G = 0.002$~$\upmu$m$^2$ps$^{-1}$, $R = 0.016$~$\upmu$m$^2$ps$^{-1}$, $\eta = 2$, $m = 0.32~\mathrm{meV}$ps${^2}$$\upmu$m$^{-2}$, and time step-size $\Delta t = 0.05$~ps.

The following simulations are single simulations using a pseudo-random initial conditions and without any stochastic elements.

\subsection*{S2 Pump profiles and placements}
In all the simulations, Gaussian pumps are used so that pump profile is of the form
\begin{align}
    P(\mathbf{r}) = P_0 \sum_{j=1}^n \exp\left( -\frac{1}{2}\left(\frac{\mathbf{r} - \mathbf{r}_j}{\sigma}\right)^2\right) 
\end{align}
where $P_0$ is the pump strength and $n$ is the number of pumps used. Throughout the simulations, $\sigma=1.27$~$\upmu$m, so the FWHM of each pump is $2\sqrt{2\ln(2)}\sigma \approx 2.99$~$\upmu$m.

To calculate the centers of the pumps a deflation algorithm~\cite{DEBRUIJN1990201} was used, by defining ten isosceles triangles arranged with one central shared vertex and the other vertices placed on a circle with a certain radius, also called the sun pattern.
The code used to simulate the system and plot Figures~\ref{fig:sup1}, \ref{fig:sup2}, \ref{fig:sup4} and \ref{fig:sup6} is available at \url{https://github.com/fixgoats/epc2dopqsm} with usage instructions.

\subsection*{S3 Results of mean field simulations}

\begin{figure} % Do not use \begin{figure*}
	\centering
	\includegraphics[width=0.9\textwidth]{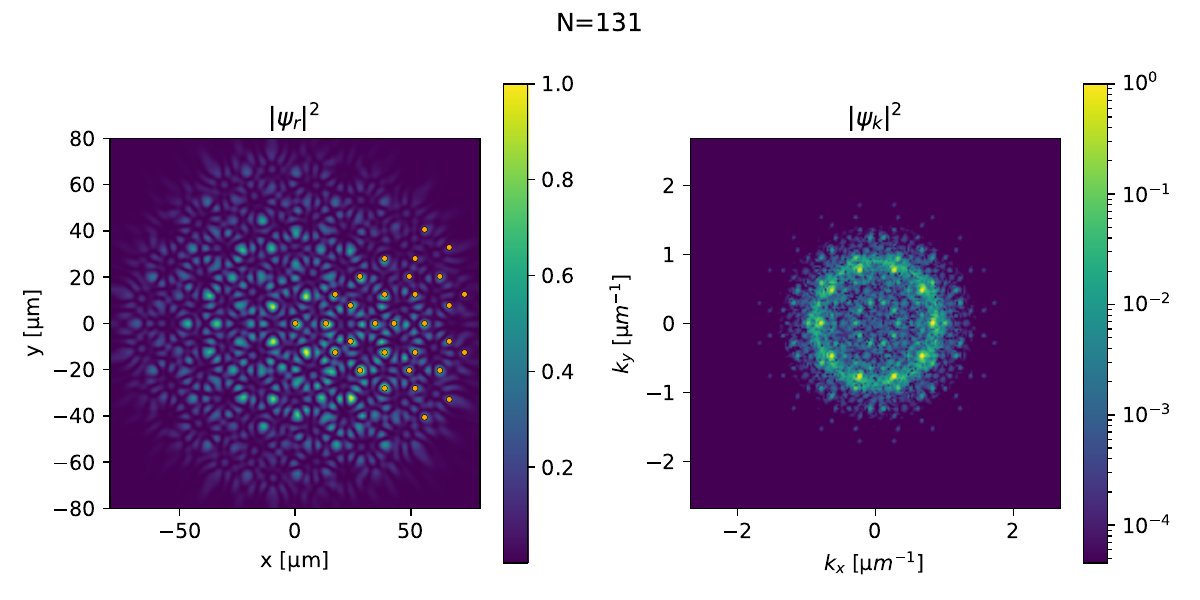} % for an image file named example_figure.*
	% Pick an appriopriate width for the size of the image

	% Captions go below figures
	\caption{\textbf{Reproduction of Figure 1 by simulation.}
		(\textbf{Left}): The $r$-space density of the condensate, normalised such that the maximum value of the density is 1. The locations of a portion of the pump spots are marked as orange dots. (\textbf{Right}): The $k$-space density of the condensate on a logarithmic scale, normalised in the same manner as the $r$-space density. The colour scale is saturated from below at $e^{-10}$. The condensation threshold is approximately $P_{th} = 7.4$~$\upmu$m$^{-2}$ and the pump strength is $P_0 = 1.4P_{th} = 10.4$~$\upmu$m$^{-2}$. The rhombi sidelengths are $D=13.2$~$\upmu$m. The brightest peaks in $\mathbf{k}$-space appear at around 0.8~$\upmu$m$^{-1}$  which is consistent with the radius of the inner ring observed experimentally.}
	\label{fig:sup1} % give each figure a logical label name
\end{figure}

In this section we demonstrate the excellent match between the steady state solutions of our condensate mean field equation~\ref{gp}. Starting from Fig.~\ref{fig:sup1}, we show there the condensate steady state corresponding to the parameters in Fig.~1 in the main manuscript. Despite the lack of translational symmetry, the simulated condensate quickly converges to a steady state solution for the given parameters and an extended ballistic wavefunction forms in real space corresponding to a set of Bragg peaks in reciprocal space. Note that in experiment we observe an additional population of higher order Bragg peaks surrounding the 10 central peaks. In simulation, the condensate population can be shifted from lower to higher order Bragg peaks by simply increasing the blueshift at each pump spot. Here, we focus on the solution dominantly occupying the lower order central Bragg peaks at a radius of $\approx 0.8$~$\upmu$m$^{-1}$ in good agreement with their location in experiment.  

\begin{figure} % Do not use \begin{figure*}
	\centering
	\includegraphics[width=0.9\textwidth]{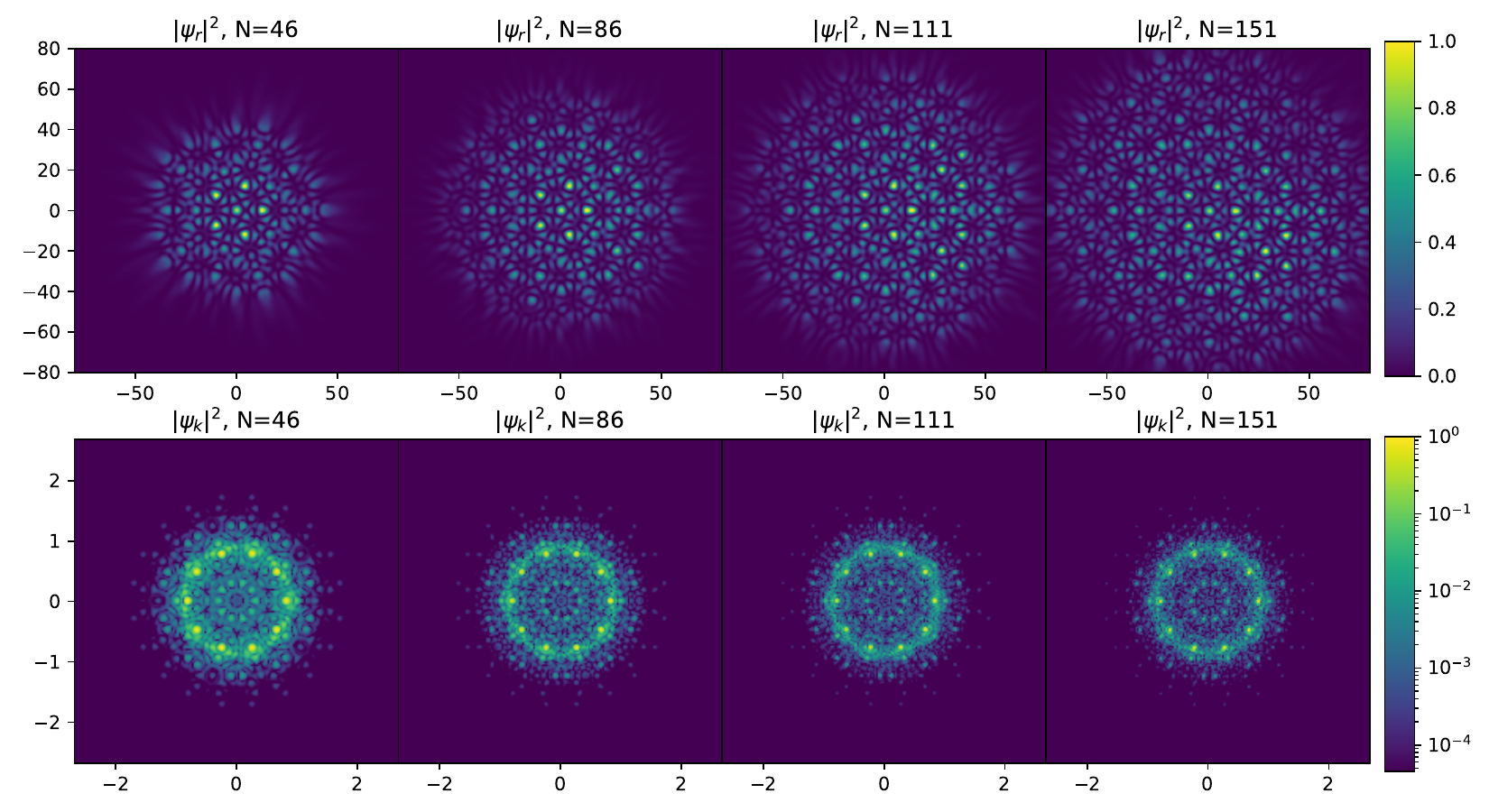} % for an image file named example_figure.*
	% Pick an appriopriate width for the size of the image

    % Captions go below figures

	\caption{\textbf{Reproduction of Figure 2 by simulation.}
		$N$ denotes the number of active optical pumps. (\textbf{Top}): $r$-space densities of the condensates, normalised so that the maximum value reached is 1. (\textbf{Bottom}): $k$-space densities of the condensates on a logarithmic scale, normalised so that the maximum value is 1. The formation of the quasicrystal is apparent by the increasing localisation of the $k$-space. The rhombi sidelengths are $D=13.2$~$\upmu$m and the pump strength is $P_0 = 10.4$~$\upmu$m$^{-2}$ in each simulation.}
	\label{fig:sup2} % give each figure a logical label name
\end{figure}

Figure~\ref{fig:sup2} shows a reproduction of Fig.~2 in the main text where we observe the occupied Bragg peaks narrowing as the lattice increases in size from left to right, in agreement with experiment. Note that our current simulations are only seeded by stochastic initial conditions and no natural broadening effects are included. The results therefore correspond to a fully coherent state (i.e., classical wavefunction). 

\begin{figure} % Do not use \begin{figure*}
	\centering
	\includegraphics[width=0.9\textwidth]{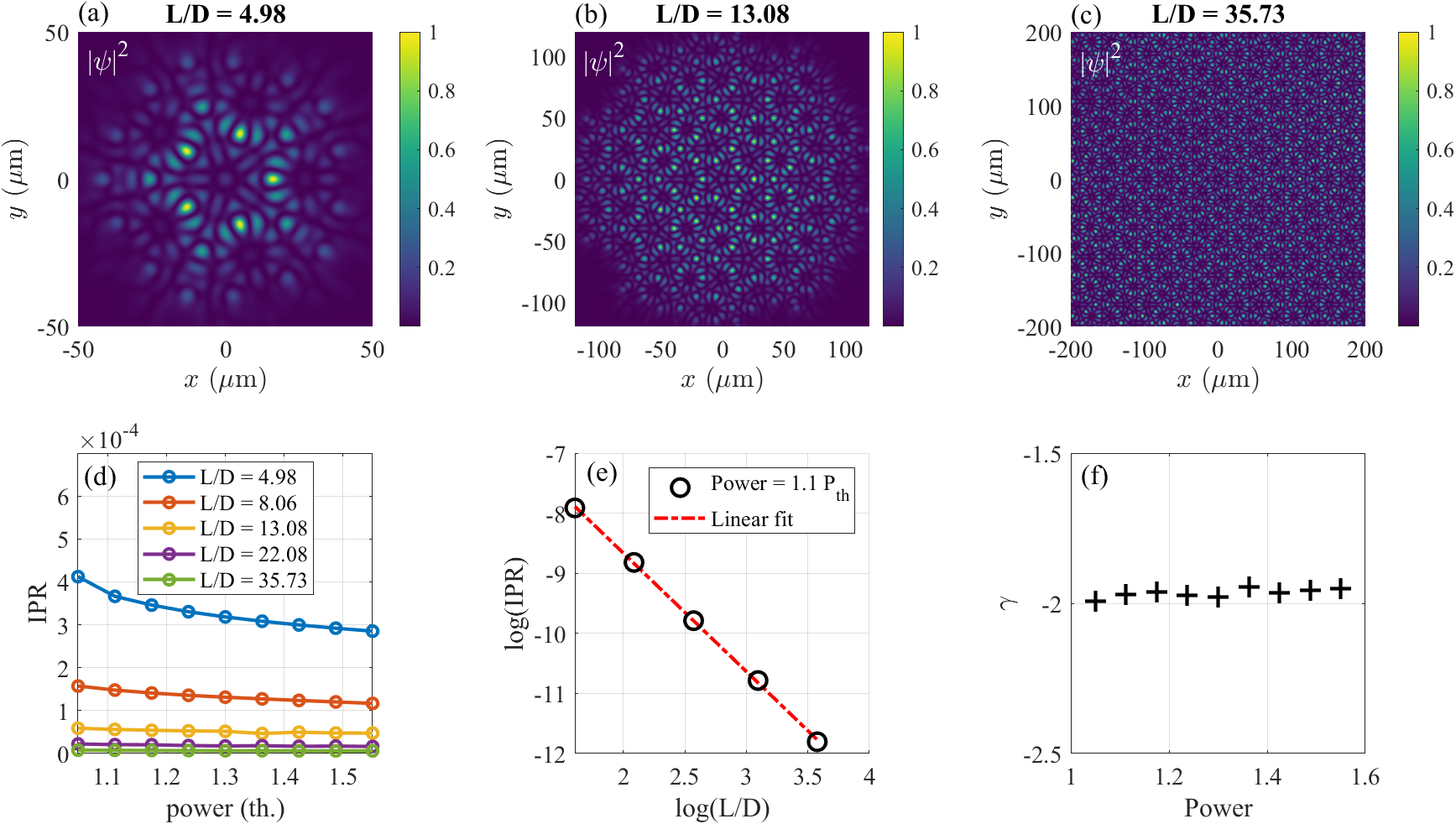} % for an image file named example_figure.*
	% Pick an appriopriate width for the size of the image

	% Captions go below figures
	\caption{\textbf{Probing localization properties of polariton quasicrystals.}
		\textbf{(a-c)} Example real space density profiles of the simulated condensate for varying number of pump spots, resulting in Penrose quasicrystals of different sizes. (\textbf{d}) Corresponding calculated IPR parameter for different system sizes and pump powers. (\textbf{e}) Logarithm of the condensate IPR for a given pump power as a function of system size resulting in a linear trend of slop $\gamma = -2$. (\textbf{f}) Approximately the same trend is observed for all other pumps powers indicating the delocalized nature of the ballistic polariton Penrose quasicrystal.}
	\label{fig:sup3} % give each figure a logical label name
\end{figure}

Figure~\ref{fig:sup3} shows a reproduction of Fig.~3 in the main text where we plot the inverse participation ratio (IPR) as a function of system size $L$ by adding more-and-more pump spots radially outwards. Here, $L$ is the distance of the furthest pump spot from the origin. Increasing the power, we observe in Fig.~\ref{fig:sup3}\textbf{d} that the IPR decreases which means that polaritons are getting expelled more strongly from their pump spots and the state is becoming more delocalized. This is in contrast to the experiment where we observe that increasing the power the IPR increases implying localization. The cause of this discrepancy is due to the fact that increased pumping causes polaritons to redshift to lower energy modes through exciton-assisted relaxation. However, plotting the IPR as a function of system size in Fig.~\ref{fig:sup3}\textbf{e} we obtain a clear linear fit with $\gamma = -2$ in full agreement with experiment, implying that our condensate steady state solutions are made up of strongly delocalized modes with long-distance coupling. The same result is obtained for all checked pump powers in as seen in Fig.~\ref{fig:sup3}\textbf{f}.

\begin{figure} % Do not use \begin{figure*}
	\centering
	\includegraphics[width=0.9\textwidth]{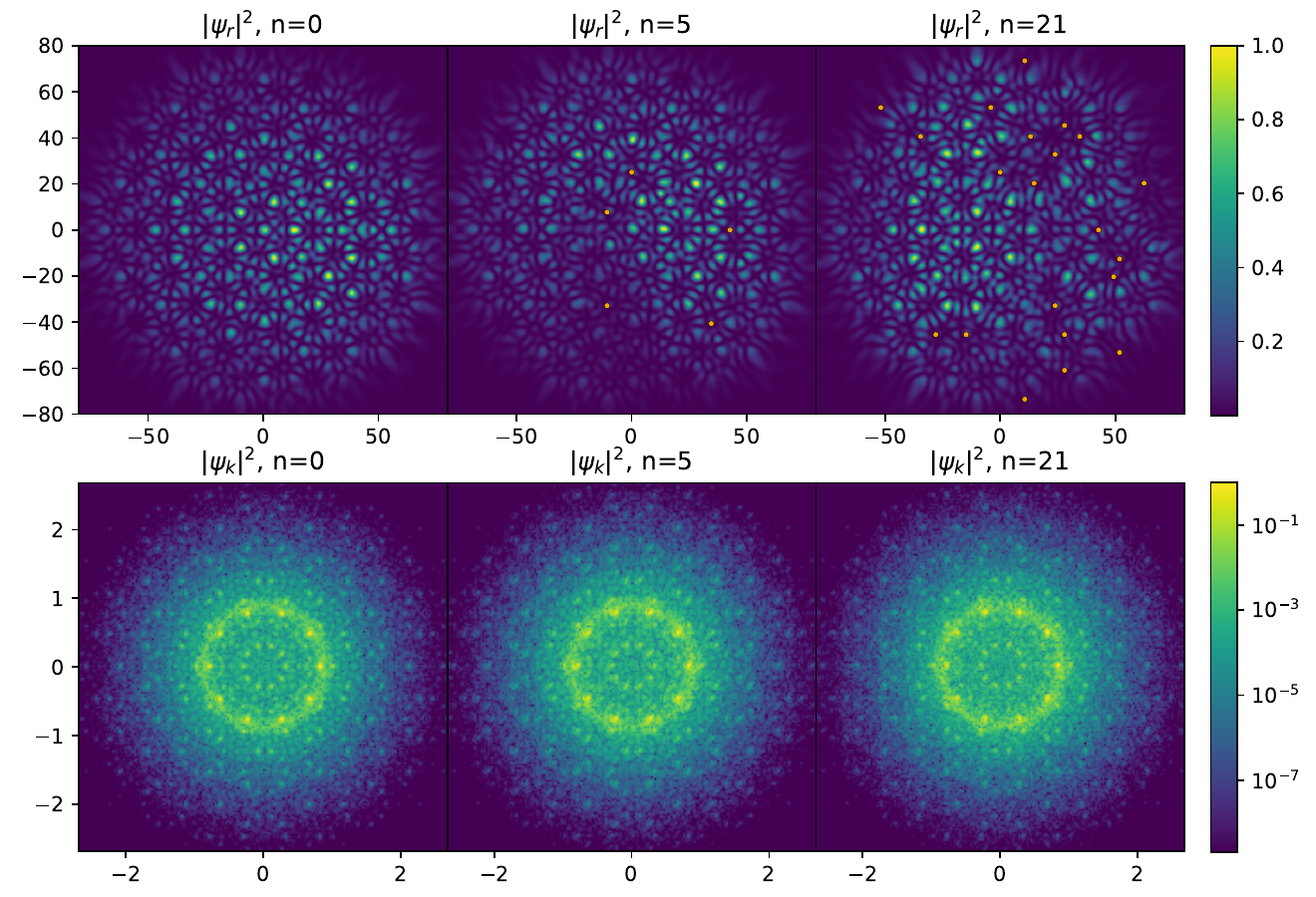} % for an image file named example_figure.*
	% Pick an appriopriate width for the size of the image

	% Captions go below figures
	\caption{\textbf{Reproduction of Figure 4 by simulation.}
		Here, $n$ is the number of defects, i.e.~randomly removed pumps, which are marked with orange dots. The decreasing definition of the quasicrystal is visually apparent at 21 defects by the blurring of the $k$-space. The rhombi sidelengths are $D=13.2$~$\upmu$m and the leftmost figure uses 131 pumps. The pumping power in each simulation is $P_0 = 10.4$~$\upmu$m$^{-2}$.}
	\label{fig:sup4} % give each figure a logical label name
\end{figure}

Figure~\ref{fig:sup4} shows a reproduction of Fig.~4 in the main manuscript where we remove by-hand 5 and then 21 randomly chosen pump spots from the lattice. Interestingly, the simulation still converges to a steady state although with skewed interference patterns in real space. The corresponding momentum space density profiles of the condensate show an increased ``fuzzyness'' as a result of this artificial disorder which manifests as increased width (uncertainty) of the populated Bragg peaks.

\begin{figure} % Do not use \begin{figure*}
	\centering
	\includegraphics[width=0.9\textwidth]{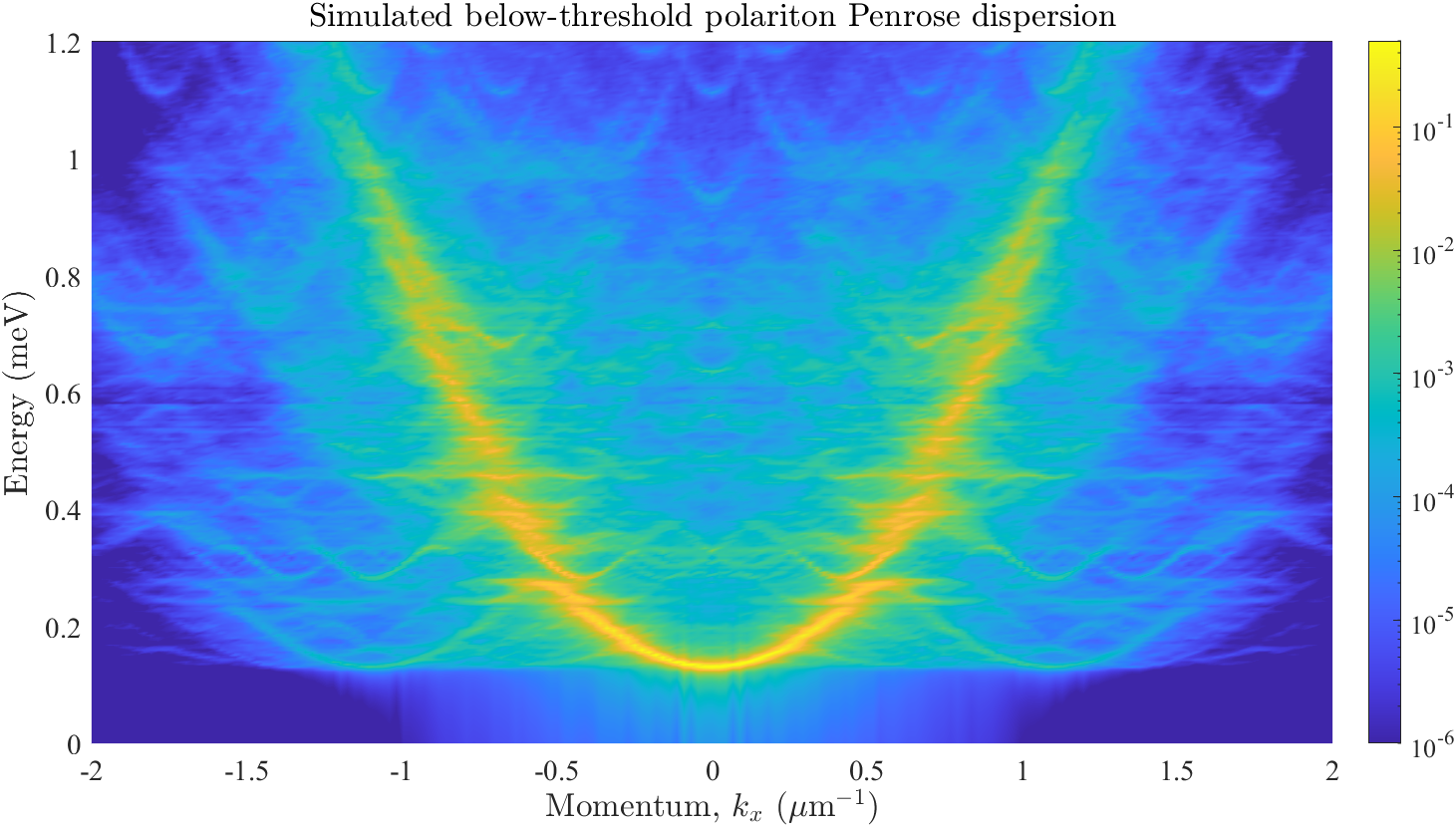} % for an image file named example_figure.*
	% Pick an appriopriate width for the size of the image

	% Captions go below figures
	\caption{\textbf{Simulated dispersion for Penrose quasicrystal.}
		Example numerically obtained dispersion along the $x$-axis for a below threshold polariton Penrose lattice (i.e., $R=0$) with a potential amplitude fixed to $v_0=2$~meV and rhombi side length $D=6$~$\upmu$m.}
	\label{fig:sup5} % give each figure a logical label name
\end{figure}

Figure~\ref{fig:sup5} shows an example dispersion below the condensation threshold. Here, in order to increase the spacing between different fractal energy branches we set the side length to $D = 6$~$\upmu$m and the peak potential amplitude of the Gaussian spots is $V_0 = 2$~meV. The results are obtained by numerically averaging the response of the finite-size Penrose lattice (with decaying boundary conditions outside the lattice perimeter) over many random white-noise initial conditions. The resulting energy-resolved Fourier space image shows a jungle of energy branches that retain sinusoidal signatures as in square and cubic lattices. However, as mentioned in the main text, the smallness of the recoil energy $E_r = \uppi^2 \hbar^2 / (2 m D^2) \approx 0.2$~meV results in multiple accessible energy manifolds for the polaritons. This result underpins the fascinating quality of the polariton condensates to synchronize and phase-lock resulting in a macroscopically coherent and single-mode condensate despite the availability many modes.

\begin{figure} % Do not use \begin{figure*}
	\centering
	\includegraphics[width=0.9\textwidth]{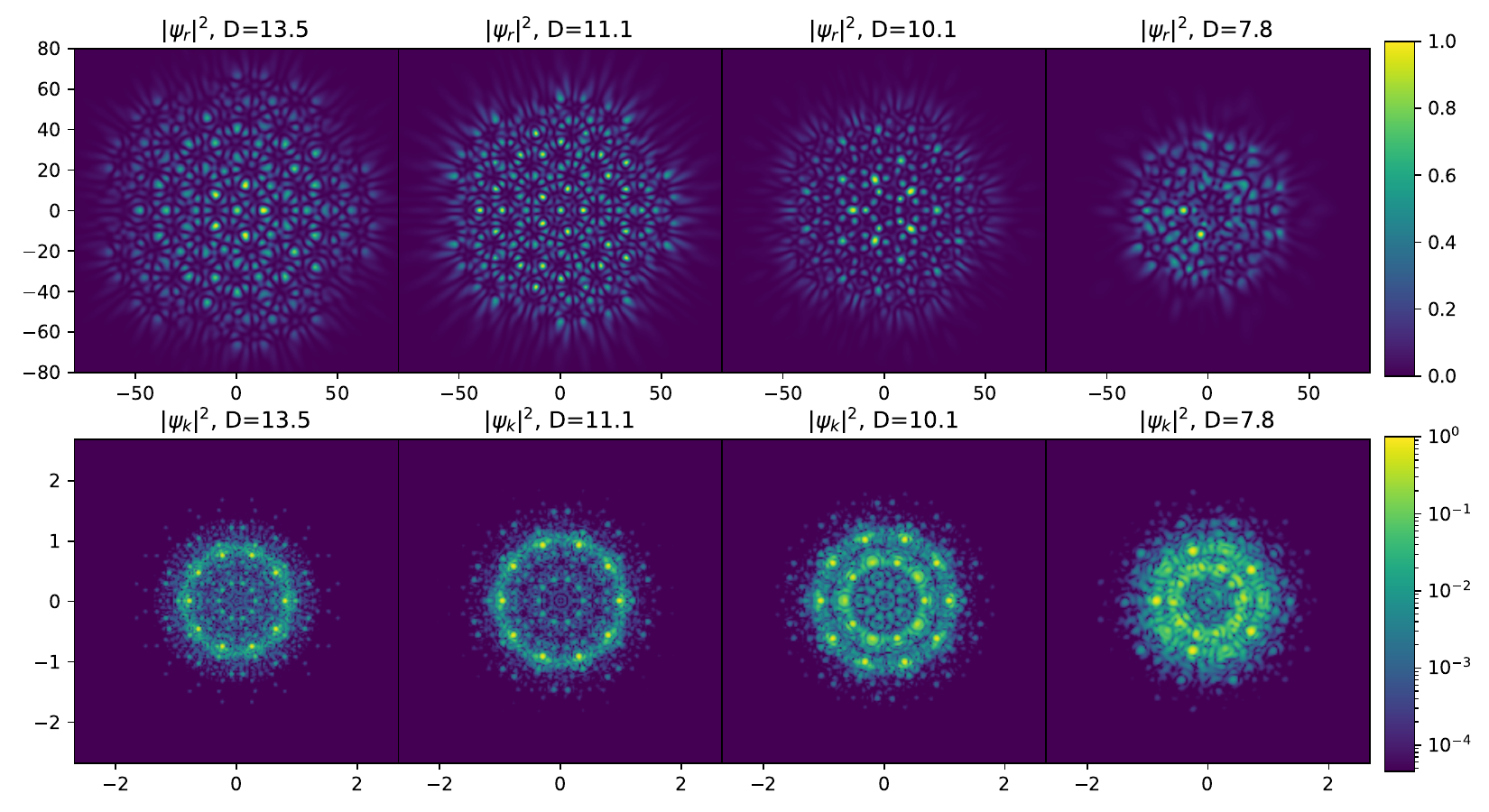} % for an image file named example_figure.*
	% Pick an appriopriate width for the size of the image

	% Captions go below figures
	\caption{\textbf{Reproduction of Figure 5.}
		$D$ denotes the rhombi sidelengths in microns, and in each figure 131 pumps are used. (\textbf{Top}) $r$-space density of the condensate, normalised so the maximum is 1. (\textbf{Bottom}) $k$-space density of the condensate, normalised so the maximum is 1. In each simulation the pumping power is $P_0 = 10.4$~$\upmu$m$^{-2}$.}
	\label{fig:sup6} % give each figure a logical label name
\end{figure}

Lastly, Fig.~\ref{fig:sup6} shows a reproduction of Fig.~5 in the main manuscript where we decrease the Penrose lattice rhombi side length (from left to right) and observe the onset of different steady state solutions. The results are in good agreement with experiment with the exception of $D = 10.1$~$\upmu$m where we do not observe the sharp Bragg peak at $k=0$ like in experiment. Our simulations do not exclude the existence of such a solution for a different set of parameters. None-the-less, our model qualitatively produce the experimental observation with the Bragg peaks losing sharpness and starting across the ballistic circle in $k$-space.

\subsection*{S4 Linear modes of quasicrystal pump landscape}

\begin{figure} % Do not use \begin{figure*}
	\centering
	\includegraphics[width=0.7\textwidth]{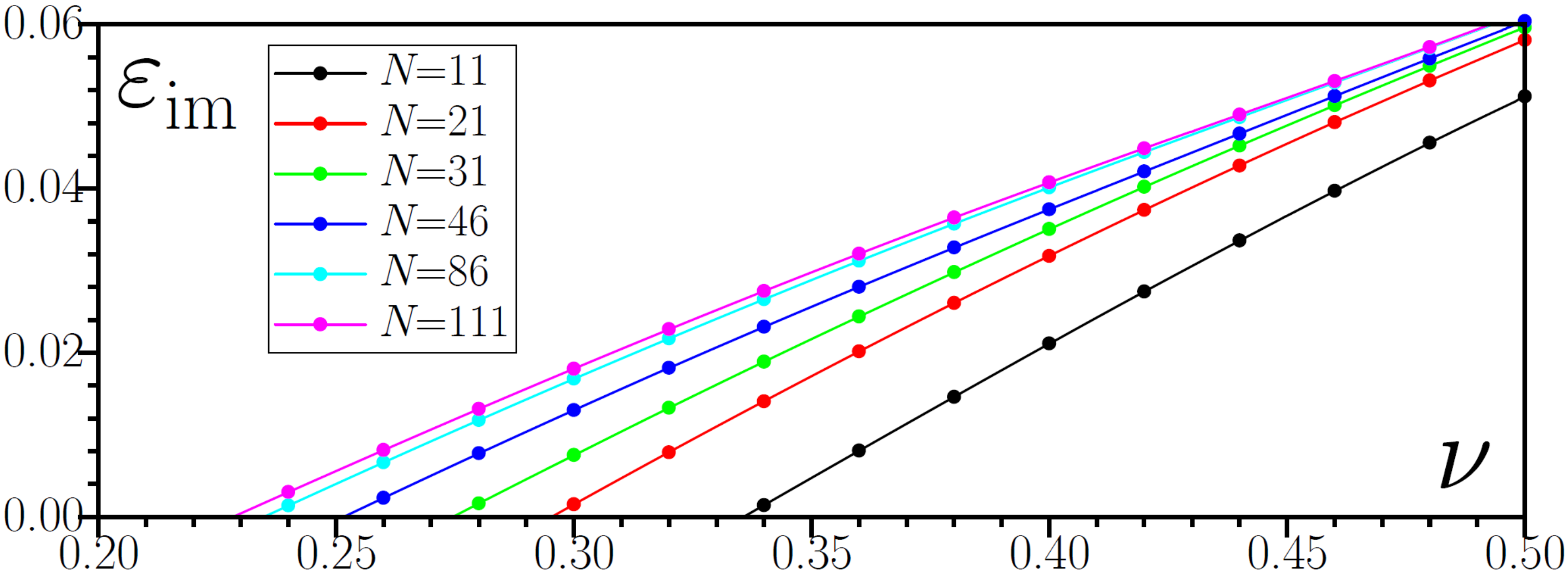} % for an image file named example_figure.*
	% Pick an appriopriate width for the size of the image

	% Captions go below figures
	\caption{\textbf{Analysis of the eigenmodes growth with pump amplitude.}
		Imaginary part of energy $\varepsilon_\textrm{im}$ of the eigenmode with fastest growth rate versus pump amplitude $\nu$ in quasicrystals with different number of nodes $N$.}
	\label{fig:sup7} % give each figure a logical label name
\end{figure}

The symmetry of the condensate right above the threshold is determined by the structure of linear modes of this dissipative system. To find such modes, we linearize the system of Eqs. (\ref{gp}) and (\ref{nr}) by omitting nonlinear terms $\sim |\psi|^2$, assuming that exciton reservoir has reached its steady state, and reducing the system to single equation for normalized polariton wavefunction $\varphi(x,y,\tau)$
\begin{equation}
    i\frac{\partial \varphi}{\partial \tau} = -\frac{1}{2} \left(\frac{\partial^2 \varphi}{\partial x^2} + \frac{\partial^2 \varphi}{\partial y^2}\right) - i\gamma \varphi + i \mathcal{I}(x,y)\varphi + \beta \mathcal{I}(x,y)\varphi \label{modes}
\end{equation}
where the coordinates $x,y$ are normalized to the characteristic scale $r_0=1$~$\upmu$m, time $\tau$ is normalized to $\hbar /\varepsilon_0$, where characteristic energy is $\varepsilon_0=\hbar^2/mr_0^2$, dimensionless loss coefficient $\gamma=\hbar \gamma_\textrm{LP}/2\varepsilon_0$, and the parameter $\beta=2g_\textrm{r}/\hbar R$. The function $\mathcal{I}(x,y)=(\hbar R/2\Gamma\varepsilon_0) P(x,y)$ describes normalized pump landscape and can be written as $\mathcal{I}(x,y)=\nu \sum_\textbf{m} e^{-[(x-x_\textbf{m})^2+(y-y_\textbf{m})^2]/\sigma^2}$, where $x_\textbf{m},y_\textbf{m}$ represent the coordinates of the quasicrystal nodes, $\nu$ is the dimensionless pump amplitude, $\sigma$ is the width of the pump spots normalized to $r_0$. For parameters of our polariton microcavity, the coefficients $\gamma\approx 0.044$ and $\beta\approx 0.695$. Here we consider the structure with rhombus side length $D=13.5$~$\upmu$m. As one can see from Eq. (\ref{modes}), the pump not only provides spatially inhomogeneous amplification, but it simultaneously creates repulsive potential $\sim \beta \mathcal{I}(x,y)$ with the same spatial structure. We search linear eigenmodes of this system in the form $\varphi(x,y,\tau)=w(x,y)e^{-i\varepsilon \tau}$, where $w(x,y)$ is the complex function describing modal shape, and $\varepsilon=\varepsilon_\textrm{re}+i\varepsilon_\textrm{im}$ is the energy that can be also complex. Substitution of wavefunction in this form into Eq. (\ref{modes}) leads to linear eigenvalue problem that we solved to find all possible eigenmodes of the system and their energies. Among them, the mode with largest $\varepsilon_\textrm{im}$ experiences preferential amplification in comparison with other modes, and therefore is likely to win the competition with other modes in the presence of nonlinear effects, determining condensate density distribution at sufficiently large evolution times. We thus determined the mode that exhibits fastest amplification and plotted imaginary part of its energy $\varepsilon_\textrm{im}$ (characterizing amplification rate) as a function of pump amplitude $\nu$ in Fig. \ref{fig:sup7} for quasicrystals with different number of nodes $N$. Pump amplitude $\nu=\nu_\textrm{th}$ at which $\varepsilon_\textrm{im}$ crosses zero allows to determine condensation threshold for a given quasicrystal configuration. One can see that this threshold decreases with increase of the number of nodes $N$ and saturates already for $N\sim 111$. $\varepsilon_\textrm{im}$ monotonically increases with pump amplitude $\nu$ for all values of $N$ presented in this figure.

\begin{figure} % Do not use \begin{figure*}
	\centering
	\includegraphics[width=\textwidth]{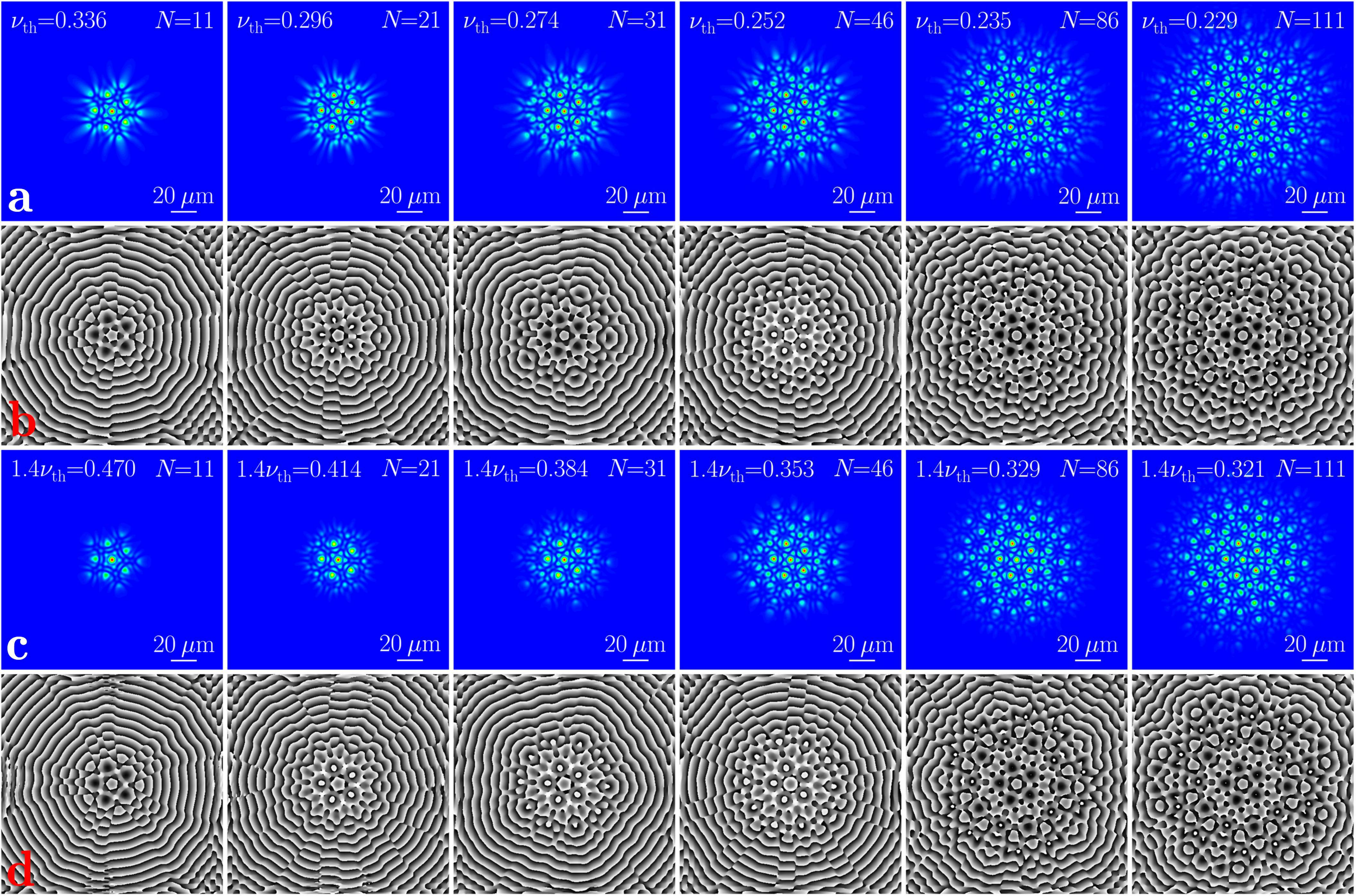} % for an image file named example_figure.*
	% Pick an appriopriate width for the size of the image

	% Captions go below figures
	\caption{\textbf{Simulated eigenmodes of polariton Penrose tiling.}
		(\textbf{a}),(\textbf{c}) Density  and (\textbf{b}),(\textbf{d}) phase  distributions in localized eigenmodes exhibiting fastest growth at threshold pump amplitude $\nu=\nu_\textrm{th}$ (\textbf{a},\textbf{b}) and above the threshold, at $\nu=1.4\nu_\textrm{th}$ (\textbf{c},\textbf{d}). Pump amplitudes are indicated on the plots. Threshold value of the pump amplitude is different for different number of nodes $N$ in quasicrystal.}
	\label{fig:sup8} % give each figure a logical label name
\end{figure}

The examples of density and phase distributions in linear eigenmodes with largest $\varepsilon_\textrm{im}$ for different values of $N$ are presented in Fig. \ref{fig:sup8} just above the condensation threshold $\nu \approx \nu_\textrm{th}$ (Fig. \ref{fig:sup8}\textbf{a},\textbf{b}) and well above this threshold at $\nu \approx 1.4\nu_\textrm{th}$ (Fig. \ref{fig:sup8}\textbf{c},\textbf{d}). Notice that all these eigenmodes correspond to positive values of $\varepsilon_\textrm{re}$. We note that away from the pumped region these eigenmodes behave as gain-guided, despite the presence of repulsive potential created by the pump. All of them are characterized by currents from the center of the mode towards the periphery, into domain where only uniform losses $\gamma$ are present. Notice that the central part of the density distribution in eigenmodes only slightly changes with increase of the number of nodes $N$ in quasicrystal structure. Eigenmodes clearly show the presence of secondary interference maxima between pumping spots, also observed in experiments. By comparing density distributions at threshold and well above the threshold, one can clearly see that the modes structure does not change significantly and only the tails outside the pumped region become less visible. It should be stressed that besides eigenmodes having the same discrete rotational symmetry as quasicrystal, the linear spectrum contains also asymmetric eigenmodes, all of which, however, have lower amplification rates $\varepsilon_\textrm{im}$ in comparison with symmetric modes depicted in Fig. \ref{fig:sup8}. Nevertheless, increasing $\nu$ leads to increase of the number of coexisting eigenmodes (symmetric and asymmetric ones) with close $\varepsilon_\textrm{im}$ values, indicating on the possibility of highly multimode nonlinear dynamics sufficiently far from the condensation threshold.

% If your supplement is very short you might need to uncomment the following line to avoid
% layout problems with the figures and tables.
%\newpage

%%%%%%%%%%%%%%%% SUPPLEMENTARY FIGURES %%%%%%%%%%%%%%%

%%%%%%%%%%% CAPTIONS FOR OTHER SUPPLEMENTARY FILES %%%%%%%%%%

\clearpage % Clear all remaining figures and tables then start a new page

%%%%%%%%%%%%%%%% SUPPLEMENTARY REFERENCES %%%%%%%%%%%%%%%

% Do NOT include a reference list in the supplement.
% All references must be in a single list at the end of the main text.
% The copyeditors will ensure that the correct reference list appears with each version of the paper
% (print, HTML, PDF, mobile app, metadata for bibliographic databases etc.)

\end{document}